\documentclass[10pt,a4paper,english,nofootinbib]{revtex4}
\usepackage{lmodern}

\usepackage[T1]{fontenc}
\usepackage[latin9]{inputenc}
\usepackage{babel}
\usepackage{amsmath}
\usepackage{amssymb}
\usepackage{esint}
\usepackage{graphicx}
\usepackage[unicode=true,pdfusetitle,
 bookmarks=true,bookmarksnumbered=false,bookmarksopen=false,
 breaklinks=false,pdfborder={0 0 1},backref=false,colorlinks=false]
 {hyperref}

\makeatletter

\pdfpageheight\paperheight
\pdfpagewidth\paperwidth

\@ifundefined{textcolor}{}
{%
 \definecolor{BLACK}{gray}{0}
 \definecolor{WHITE}{gray}{1}
 \definecolor{RED}{rgb}{1,0,0}
 \definecolor{GREEN}{rgb}{0,1,0}
 \definecolor{BLUE}{rgb}{0,0,1}
 \definecolor{CYAN}{cmyk}{1,0,0,0}
 \definecolor{MAGENTA}{cmyk}{0,1,0,0}
 \definecolor{YELLOW}{cmyk}{0,0,1,0}
 }

\usepackage{latexsym}\usepackage{bm}

\makeatother

\begin{document}
\title{Gravitomagnetism in Massive Gravity}

\author{Kezban Tasseten}

\email{tasseten@metu.edu.tr}

\affiliation{Department of Physics,\\
 Middle East Technical University, 06800, Ankara, Turkey}

\author{Bayram Tekin}

\email{btekin@metu.edu.tr}

\affiliation{Department of Physics,\\
 Middle East Technical University, 06800, Ankara, Turkey}
\date{\today}

\begin{abstract}

Massive gravity in the weak field limit  is described by the Fierz-Pauli theory with 5 degrees of freedom in four dimensions. In this theory, we calculate the gravitomagnetic effects (potential energy) between two point-like, spinning sources that also orbit around each other in the limit where the spins and the velocities are small. Spin-spin, spin-orbit and orbit-orbit interactions in massive gravity theory have rather remarkable, discrete differences from their counterparts in General Relativity. Our computation is applicable for large distances, for example, for interaction between galaxies or galaxy clusters where massive gravity is expected to play a role. We also extend the computations to quadratic gravity theories in four dimensions and find the lowest order gravitomagnetic effects and show that at small separations quadratic gravity behaves differently than General Relativity.

\end{abstract}
\maketitle

{\section {Introduction}}

 One of the simplest modifications of General Relativity (GR) with improved infrared (IR) behavior that could possibly explain the observed accelerated expansion of the Universe is Massive Gravity (mGR), a topic which has been around since the work of Fierz and Pauli in 1939 \cite{Fierz_Pauli} but received a revived interest in the past decade for various   reasons and in various forms (See the excellent reviews on the subject \cite{Hinterbichler, deRham}). At the lowest order, in four dimensions, the Lagrangian density is that of the Fierz-Pauli (FP) theory 
\begin{align}
\mathcal{L}_{mGR}  =\frac{1}{16\pi G}\left[R-\frac{m_{g}^{2}}{4}\left(h_{\mu\nu}^{2}-h^{2}\right)\right]
+{\cal L}_{\text{matter}},
\end{align}
which propagates a non-ghost, non-tachyonic massive (with mass $m_{g}$) spin-2 particle with all 5 helicity modes in flat backgrounds. \emph{By construction}, at the level of the Lagrangian, or at the level of the field equations, as $m_{g}$ $\rightarrow 0$, GR is smoothly recovered. But it is well-known that once the Newtonian potential is computed between two static sources or when deflection of light is computed as it passes a static source, one of these two results can be matched to the GR value by redefining the Newton's constant but the other one does not smoothly reproduce the Newtonian (or the GR) limit. For example the potential energy in  mGR between  two static sources reads 
\begin{equation}
U_{mGR}=-\dfrac{4}{3} \dfrac{Gm_{1}m_{2}}{r} e^{-m_{g} \, r},
\end{equation}
and as $m_{g}$ $\rightarrow 0$, one has the famous van Dam-Veltman-Zakharov (vdVZ) \cite{DV,Zak} discontinuity  which has been known since early 1970s.
All this is well established in the literature. But, what is rather interesting is that, only recently \cite{Gullu_Tekin1}, another discontinuity between GR and mGR was found in the spin orientations of spinning point masses (such as two widely separated stars, galaxies or galaxy clusters). In GR, the spin-spin interaction (when other interactions such as tidal forces etc are neglected) can be computed as 
\begin{equation}
U^{\mbox{spin-spin}}_{GR}=-\frac{G}{r^{3}}\left[\vec{J_{1}}\centerdot\vec{J_{2}}-3\vec{J_{1}}\centerdot\hat{r}\,
\vec{J_{2}}\centerdot\hat{r}\right],\label{general_relativity}
\end{equation}
 where $\vec{J_{1}}$ and $\vec{J_{2}}$ are the spins  of the  sources and $\vec{r}$ is the radial vector between them. (Here, spin refers to  rotation about an axis passing through the object which we approximately take as point-like.). 
\begin{figure}[h]
\includegraphics[width=15 cm]{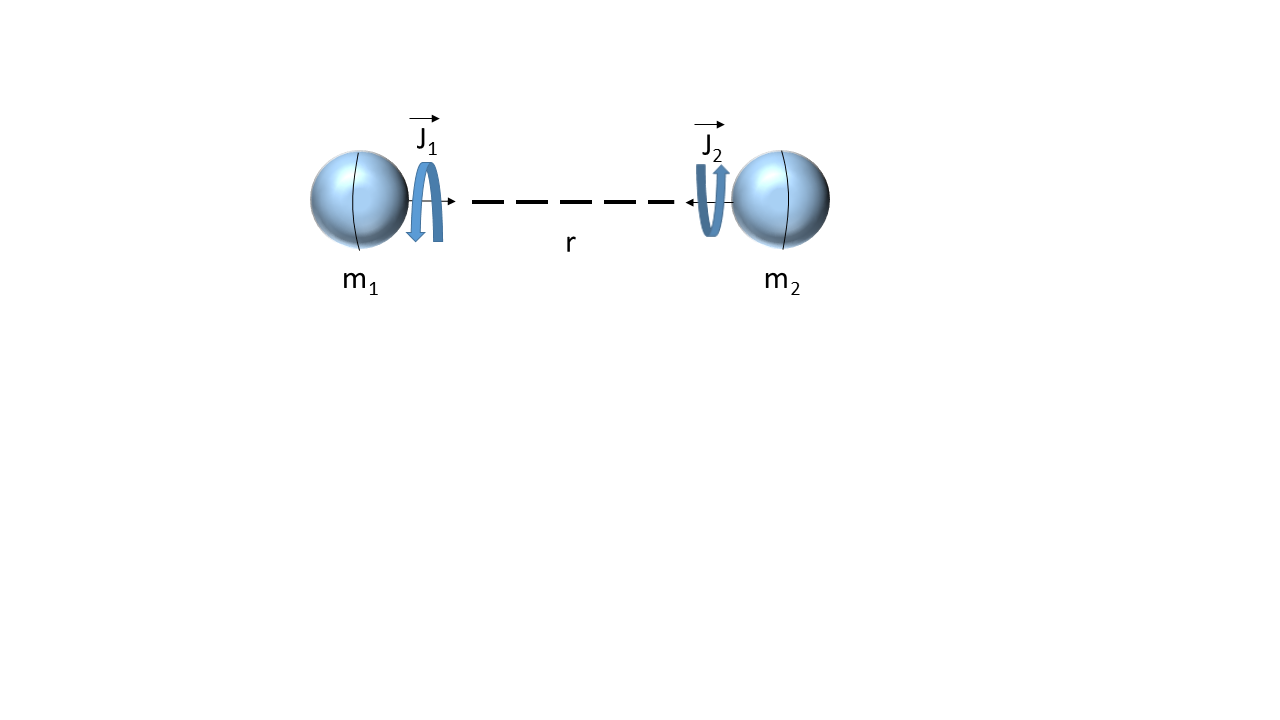}\caption{Minimum energy configuration in GR, as long as weak field limit is applicable: spins are anti-parallel to each other, so the spin is minimized.}
\end{figure}
 On the other hand,  in mGR, the same quantity can be found as \cite{Gullu_Tekin1}
\begin{align}
U^{\mbox{spin-spin}}_{mGR}=  -\frac{Ge^{-m_{g}r}\left(1+m_{g}r
	+m_{g}^{2}r^{2}\right)}{r^{3}}\left[\vec{J_{1}}\centerdot\vec{J_{2}}-3\vec{J_{1}}\centerdot\hat{r}\,
\vec{J_{2}}\centerdot\hat{r}\frac{\left(1+m_{g}r+\frac{1}{3}m_{g}^{2}r^{2}\right)}{1+m_{g}r+m_{g}^{2}r^{2}} 
\right], \label{mGR}
\end{align}
\begin{figure}[h]
\includegraphics[width=15 cm]{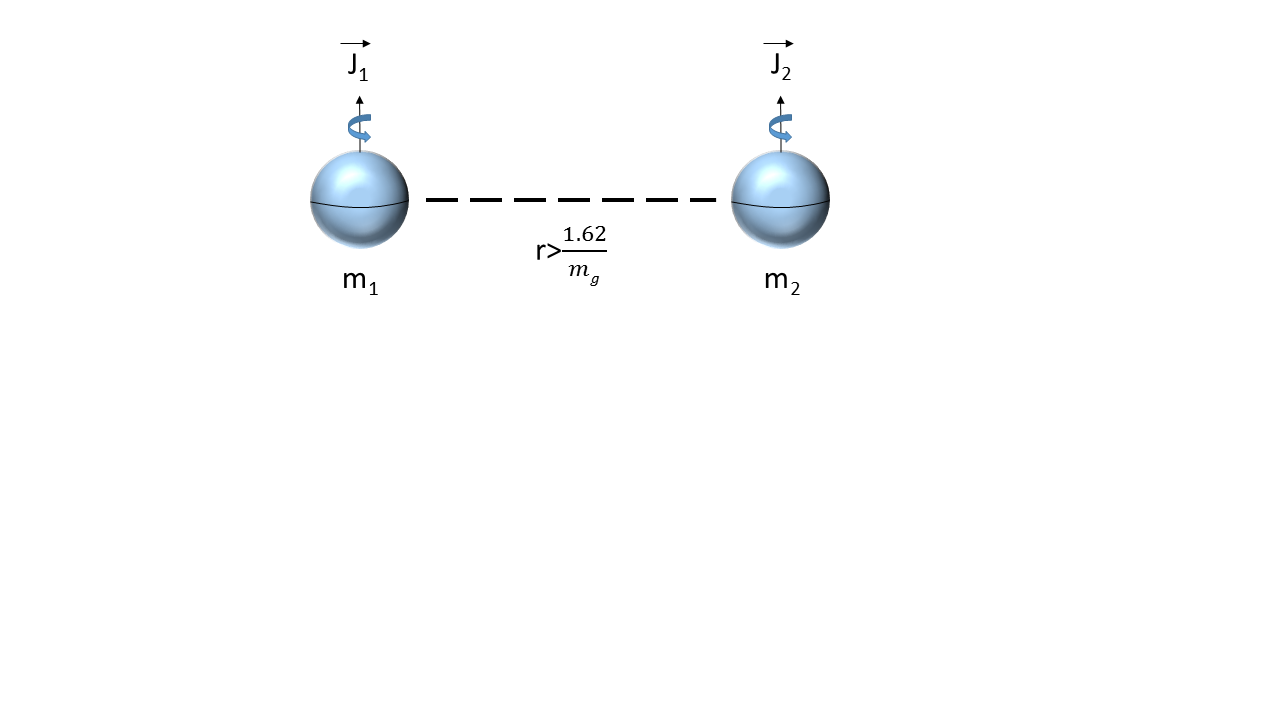}\caption{Minimum energy configuration in massive gravity: for distances $r>1.62 / m_{g}$, spins are parallel and point in the direction perpendicular to the axis joining the sources. For small separations mGR reduces to GR.}
\end{figure}
Observe that as $m_{g}$ $\rightarrow 0$, exponential decay part is expected since massive gravity is weaker at large distances and even perhaps, one could expect, just like the Newtonian potential case, a discrete numerical difference in the overall factor of the potential. But, what is highly surprising is the discrete numerical difference between the two spin-spin terms in ({\ref{mGR}}): As  $r$ $\rightarrow \infty$, one has 

\begin{align}
U^{\mbox{spin-spin}}_{mGR}\longrightarrow -\frac{G}{r^{3}}\left[\vec{J_{1}}\centerdot\vec{J_{2}}-\vec{J_{1}}\centerdot\hat{r}\,
\vec{J_{2}}\centerdot\hat{r}\right],
\end{align}
which differs discretely from that of GR ({\ref{general_relativity}}). But that is \emph{not} the whole surprise : As we expect from the Newtonian attraction such a discrete discontinuity might arise. The real surprise is that this numerical difference leads to different observable spin orientations in these two theories. While GR favors anti-parallel spins ( as shown in Figure 1), minimizing the total spin of the system; massive GR favors parallel spins perpendicular to the line joining the sources, maximizing the total spin of the system ( as shown in Figure 2) . The fact that a tiny mass leads to such a remarkable effect is rather amazing. A detailed analysis in \cite{Gullu_Tekin1}, actually shows that for a given graviton mass $m_{g}$,
for distances that satisfy $m_{g}\, r \geq \dfrac{1+\sqrt{5}}{2}\approx 1.62$  the spin of the system is maximized as shown in Figure 2. On the other hand, for distances that satisfy  $m_{g}\, r < \dfrac{1+\sqrt{5}}{2}$, massive gravity agrees with GR and spin of the system is minimized. At the point $m_{g}\, r_{*} \approx 1.62$, the relative coefficient between the two terms in ({\ref{mGR}}) becomes $-2$.  Therefore, as expected at the large separations in the Universe, mGR can show its effects on the orientations of galaxies or galaxy clusters. In the Conclusion section, we will point out to some possible observations along these lines. 

Encouraged by these interesting results, in this work, we continue our analysis on the other and higher order gravitomagnetic effects in massive gravity as well as higher derivative gravity theories.
We shall assume that the two spinning sources also have velocities, hence orbital motion which will yield  additional effects that could probably lead to interesting differences between GR and mGR. 

Our computations in quadratic gravity are in some sense academic in nature since the differences will arise in UV not in IR:  At large separations, clearly one expects GR to dominate over the quadratic terms, but at small separations, quadratic terms are more effective. 

Let us say a few words about how we shall proceed to compute the promised effects. The "canonical" way of finding gravitomagnetic effects in a given gravity theory at the weak field, small velocity, small spin limit is to find the linearized Kerr solution of the theory and to compute the "motion" of a spinning, orbiting test particle in this geometry. But this is a rather long and cumbersome procedure, instead of that we shall employ a more clean-cut method, that is, we will compute the tree-level  scattering amplitude due to one graviton exchange between two covariantly conserved sources. Details of this procedure has been laid out in \cite{Gullu_Tekin2} but it pays to recapitulate here for the sake of completeness. 

The lay-out of the paper is as follows: In section II, and III we show how the gravitomagnetic potential energy including all the terms up to desired order can be computed from the tree-level scattering amplitude. In section IV, which is the bulk of the paper, we give a detailed derivation of the potential energy for the moving, spinning masses in both GR and mGR. Some of the details of the computations are relegated to the Appendix. In section V, we calculate the analogous expressions for quadratic gravity (with no explicit Fierz-Pauli mass term).

\section{Potential energy from  graviton exchange:tree-level scattering }

To relate the potential energy at the desired order to the scattering amplitude, let us follow \cite{zee} and compute the vacuum to vacuum transition amplitude in the path integral formalism between two sources: 

\begin{equation}
\langle 0 | e^{-i H t} | 0\rangle =  e^{-i U t} = W[T] = \int {\cal{D}}\, h_{\mu \nu}  e^{i S [h, T]} , 
\end{equation}
where $t$ is a large time that will drop at the end and  $S[h, T]$ is the linearized action about a background ($\bar{g}_{\mu \nu})$ in a generic gravity model which reads 
\begin{equation}
S[h, T]= \int d^D x  \sqrt{-\bar{g}} \Big \{ - \frac{1}{ \kappa} h^{\mu \nu} {\cal{ E}}_{\mu \nu}(h)+  h^{\mu \nu} T_{\mu \nu} \Big \}.
\end{equation}
From this follow the linearized field equations
\begin{equation} 
{\cal{ E}}_{\mu \nu}(h) = \frac{\kappa}{2} T_{\mu \nu}. \label{field_eqn}
\end{equation}
Covariant conservation of $T_{\mu \nu}$ leads to   $ \bar{\nabla}_\mu {\cal{ E}}^{\mu \nu}(h) =0$. Let us employ the background field method to recast the path-integral in a more manageable form. For this purpose, suppose $\bar{h}_{\mu \nu}$ satisfies (\ref{field_eqn}), then make a change of variables in the path-integral as   $h_{\mu \nu } \rightarrow 
h_{\mu \nu} + \bar{h}_{\mu \nu}$ which  does not change the measure, but shifts the action to the decoupled form 
\begin{equation}
S[h, T]= \int d^D x  \sqrt{-\bar{g}} \Big \{ - \frac{1}{ \kappa } h^{\mu \nu} {\cal{ E}}_{\mu \nu}(h)+\frac{1}{2} \bar{h}^{\mu \nu} T_{\mu \nu} \Big \}.
\end{equation}
The second term is free of $h_{\mu \nu }$  so we can move it out of the path-integral and the first term, being independent of $T_{\mu \nu}$, simply rescales the normalization factor, yielding   
\begin{equation}
 W[T] = {\cal{N}} e^{\frac{ i}{2}\int d^D x \sqrt{-\bar{g}}\, \bar{h}^{\mu \nu} T_{\mu \nu} }.
\end{equation} 
Equation (\ref{field_eqn}) is of the form
\begin{equation}
{\cal{O}}_{\mu \nu \alpha \beta}(x) \bar{h}^{\alpha \beta}(x) = \frac{\kappa}{2} T^{\alpha  \beta}(x),
\end{equation}
where  ${\cal{O}}_{\mu \nu \alpha \beta}$ is a self-adjoint operator whose Green's function is defined as 

\begin{equation}
{\cal{O}}_{\mu \nu \alpha \beta} G^{\alpha \beta}\,_{\sigma \rho}(x,x') = \frac{1}{2} \Big ( \bar{g}_{\mu \sigma} \bar{g}_{\nu \rho} +\bar{g}_{\mu \rho} \bar{g}_{\nu \sigma} \Big ) \delta(x-x').
\end{equation}
Therefore the particular solution of (\ref{field_eqn}) can be formally written as 

\begin{equation}
\bar{h}_{\mu \nu}(x) = \frac{\kappa}{2} \int d^D x'  \sqrt{-\bar{g}}\, G_{\mu \nu \alpha \beta }(x, x') T^{\alpha \beta}(x').
\end{equation}
Therefore one has the usual source-source interaction 
\begin{equation}
 W[T] = {\cal{N}} e^{\frac{i \kappa} {4}\int d^{D}x\,\int d^{D}x'\sqrt{-\bar{g}(x)}\sqrt{-\bar{g}(x')} T^{\mu\nu}\left(x\right)G_{\mu\nu\alpha\beta}\left(x,x^{\prime}\right)T^{\alpha\beta}\left(x^{\prime}\right) },
\end{equation}
from which we can read the potential energy (up to an irrelevant constant) as
\begin{equation}
U= -{\frac{ \kappa} {4 t }\int d^{D}x\, d^{D}x'\sqrt{-\bar{g}(x)}\sqrt{-\bar{g}(x')} T^{\mu\nu}\left(x\right)G_{\mu\nu\alpha\beta}\left(x,x^{\prime}\right)T^{\alpha\beta}\left(x^{\prime}\right) }. \label{pot}
\end{equation}
 We have kept the discussion to the $D$-dimensional, for $D=3+1$, we have   $\kappa = 16 \pi G$.  Figure 3 represents the interaction.
\begin{figure}
	\includegraphics{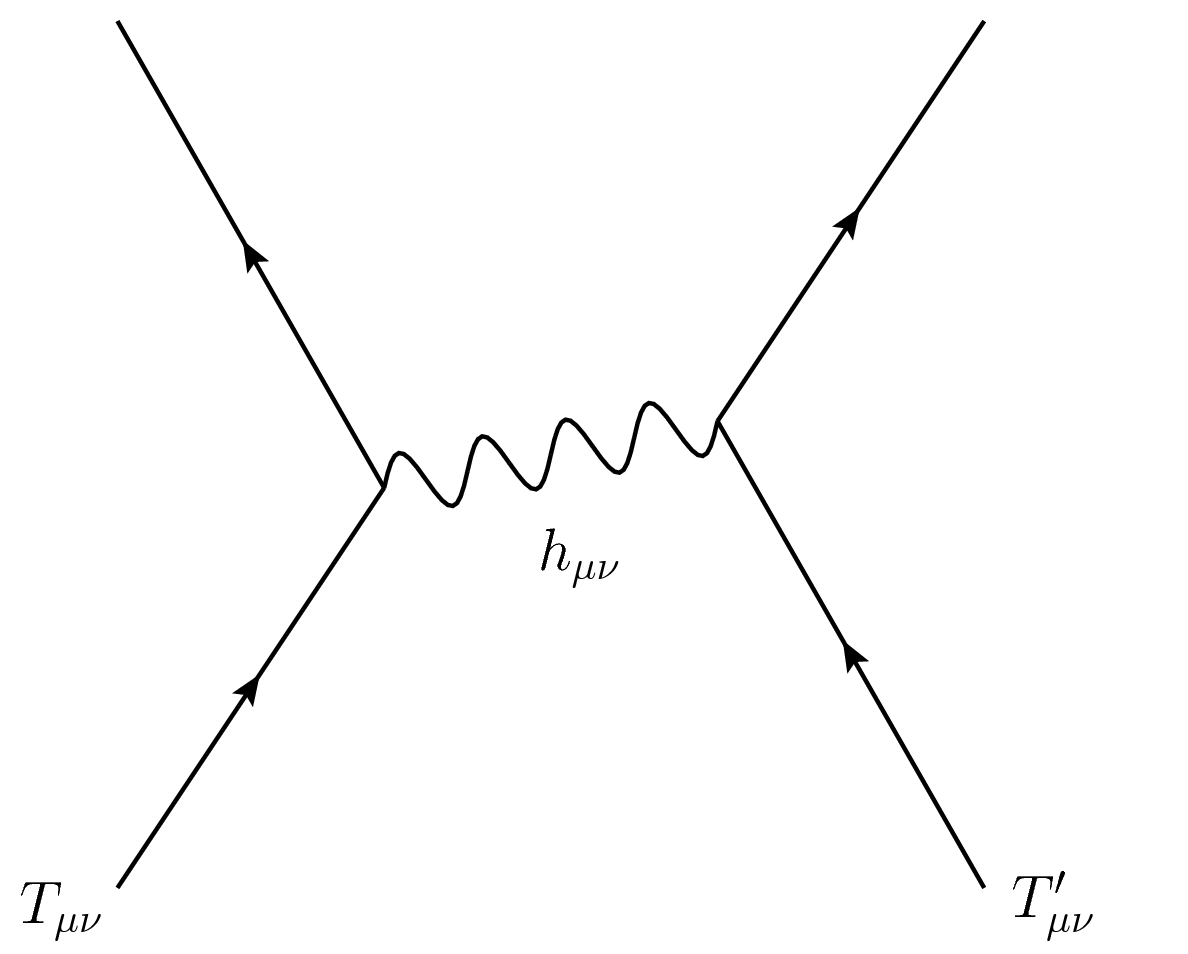}\caption{Tree level scattering digram, between two conserved sources. From this diagram, we compute the potential energy.}
\end{figure}
\section{Computation of the potential energy }

We shall compute the mentioned gravitomagnetic effects in three different theories: General Relativity, (linearized) massive gravity and quadratic gravity. To be able to employ (\ref{pot}) we need the propagators of these theories. All these theories have different propagators which require a lot of work to carry out the full computation. To reduce the amount of computation, we shall use the "master" scattering amplitude that was found in \cite{Gullu_Tekin2}, which upon taking the corresponding limits, generates the relevant scattering amplitudes in these three theories. There is a caveat though, to be able to smoothly reproduce the GR's and mGR's scattering amplitudes from one single expression, one must introduce o provisional cosmological constant which is set to zero before $m_{g}\longrightarrow 0 $ \cite{kogan},\cite{porrati}. The relevant action is, 
\begin{eqnarray}
S & = & \int d^{D}x\,\sqrt{-g}\left\{ \frac{1}{\kappa}R-\frac{2\Lambda_{0}}{\kappa}+\alpha R^{2}+\beta R_{\mu\nu}^{^{2}}+\gamma\left(R_{\mu\nu\sigma\rho}^{2}-4R_{\mu\nu}^{2}+R^{2}\right)\right\} \nonumber \\
 &  & +\int d^{D}x\,\sqrt{-g}\left\{ -\frac{m_g^{2}}{4\kappa}\left(h_{\mu\nu}^{2}-h^{2}\right)+{\cal L}_{\text{matter}}\right\} ,\label{action}
\end{eqnarray}
and the tree-level one graviton exchange in this theory is \cite{Gullu_Tekin2} 
\begin{eqnarray}
4Ut & = & 2\int d^{D}x\,\sqrt{-\bar{g}}\left\{ T{}_{\mu\nu}^{\prime}\left(x^{\prime}\right)\left\{ (\beta\bar{\square}+a)(\triangle_{L}^{(2)}-\frac{4\Lambda}{D-2})+\frac{m_g^{2}}{\kappa}\right\} ^{-1}T^{\mu\nu}\left(x\right)\right\} \nonumber \\
 & + & \frac{2}{D-1}T^{\prime}\left\{ (\beta\bar{\square}+a)(\bar{\square}+\frac{4\Lambda}{D-2})-\frac{m_g^{2}}{\kappa}\right\} ^{-1}T\label{mainresult}\\
 & - & \frac{4\Lambda}{(D-2)(D-1)^{2}}T^{\prime}\left\{ (\beta\bar{\square}+a)(\bar{\square}+\frac{4\Lambda}{D-2})-\frac{m_g^{2}}{\kappa}\right\} ^{-1}\left\{ \bar{\square}+\frac{2\Lambda D}{(D-2)(D-1)}\right\} ^{-1}T\nonumber \\
 & + & \frac{2}{(D-2)(D-1)}T^{\prime}\left\{ \frac{1}{\kappa}+4\Lambda f-c\bar{\square}-\frac{m_g^{2}}{2\kappa\Lambda}(D-1)\right\} ^{-1}\left\{ \bar{\square}+\frac{2\Lambda D}{(D-2)(D-1)}\right\} ^{-1}T,\nonumber 
\end{eqnarray}
which looks rather cumbersome, but bear in mind that it encompasses all the information one needs for the theories that we are interested in. $\bar{\square}$ is the d'Alembertian in the $\bar{g}_{\mu \nu}$ background. The parameters that appear in this scattering amplitude are defined as  
\begin{eqnarray}
\nonumber
&f&\equiv\left(D\alpha+\beta\right)\frac{\left(D-4\right)}{\left(D-2\right)^{2}}+\gamma\frac{\left(D-3\right)\left(D-4\right)}{\left(D-1\right)\left(D-2\right)} , \\
& a& \equiv\frac{1}{\kappa}+\frac{4\Lambda D}{D-2}\alpha+\frac{4\Lambda}{D-1}\beta+\frac{4\Lambda\left(D-3\right)\left(D-4\right)}{\left(D-1\right)\left(D-2\right)}\gamma, \nonumber \\
& c& =  \frac{4(D-1)\alpha+ D\beta }{D-2}, 
\end{eqnarray}
where the effective cosmological constants can be found from $\frac{\Lambda-\Lambda_{0}}{2\kappa}+f\Lambda^{2}=0\label{quadratic}$. For GR, in flat space the proper limits are 
$\dfrac{m_g^2}{\Lambda}\longrightarrow 0$,  $\alpha=\beta=\gamma=0$,  $a=\frac{1}{\kappa}$ which yield 
\begin{equation}
2Ut =- \kappa T{}_{\mu\nu}^{\prime}(\partial^{2}) ^{-1}T^{\mu\nu}+\frac{ \kappa}{D-2} T^{\prime}(\partial^{2}) ^{-1}T,  \\
\end{equation}
where we have suppressed the integral signs, but more explicitly one has 
\begin{equation}
2Ut= -\kappa\int d^{D}x\int d^{D}x^{\prime}T_{\mu\nu}\left(x^{\prime}\right)G\left(x,x^{\prime}\right)T^{\mu\nu}\left(x\right)+\frac{\kappa}{\left(D-2\right)}\int d^{D}x\int d^{D}x^{\prime}T\left(x^{\prime}\right)G\left(x,x^{\prime}\right)T\left(x\right),
\end{equation}
where we have employed the scalar Green's function which is easier to handle 
\begin{equation}
\partial_{x}^{2}G\left(x,x^{\prime}\right)=-\delta\left(x,x^{\prime}\right),
\end{equation}
with the flat space d'Alembertian  $\partial_{x}^{2}=-\partial_{t}^{2}+\vec{\nabla}^{2}$.  In the explicit computations below,  we use the retarded Green's function which are easily found as
\begin{equation}
\left(\partial^{2}\right)^{-1}\equiv G_{R}\left(x,\, x^{\prime}\right) = \frac{\Gamma\left(\frac{D-3}{2}\right)}{4\pi^{\frac{D-1}{2}}r^{D-3}}\delta\left[r-\left(t-t^{\prime}\right)\right],\label{laplacian_green}
\end{equation}
for the massless case and 
\begin{equation}
\left(\partial^{2}-m_{g}^{2}\right)^{-1}\equiv  G_{R}\left(x,\, x^{\prime}\right)=\frac{\left(\frac{m_{g}}{r}\right)^{\frac{D-3}{2}}}{\left(2\pi\right)^{\frac{D-1}{2}}}K_{\frac{D-3}{2}}\left(r\, m_{g}\right)\delta\left[r-\left(t-t^{\prime}\right)\right],\label{massive_greens}
\end{equation}
for the massive case. Here $K_{\nu}$ is the modified Bessel function of the second kind. Up to this point we have kept the discussion to be in generic $D$ dimensions, but from now on we shall stick to the $D=3+1$ dimensional case. In \cite{Gullu_Tekin1}, spin-spin interaction was found in $D$ dimensions but because the spin is a higher rank tensor in more than 4 dimensions, the expressions become quite non-trivial, even if the computations are straightforward. 

Before we close this section, let us digress a little bit and compute the interaction of two massless sources (say photons) moving at the speed of light, the ultra relativistic limit that we shall avoid in the rest of this work, and reproduce the old, amusing result of \cite{tolman}, that two anti-parallel moving light beams attract each other, while two parallel moving beams do not interact at all (note that \cite{zee} has a nice discussion on this issue). Let the four velocities be  
\begin{align}
u^{\mu}_{1}=(1,0,0,1) , \hspace{1 cm} u^{\mu}_{2}=(1,0,0,\sigma) ,
\end{align}
where $\sigma=1$ refers to parallel motion and $\sigma=-1$ anti-parallel motion. The energy-momentum tensor for each photon is
\begin{align}
T^{\mu \nu} =Eu^{\mu}u^{\nu},
\end{align}
with $ T=0$. Vanishing of the trace  immediately says that the result will be the same in GR and mGR. Then one has, 
\begin{align}
4Ut&= -2\kappa T{}_{\mu \nu}^{\prime} \left( \partial^{2}\right) ^{-1}T{}^{\mu \nu}, \nonumber \\
&=-2\kappa T{}_{00}^{\prime} \left( \partial^{2}\right) ^{-1}T{}^{00}-4\kappa T{}_{0i}^{\prime} \left( \partial^{2}\right) ^{-1}T{}^{0i}-2\kappa T{}_{ij}^{\prime} \left( \partial^{2}\right) ^{-1}T{}^{ij}.
\end{align}
Upon carrying out the integrals one arrives at
\begin{align}
T{}_{00}^{\prime} \left( \partial^{2}\right) ^{-1}T{}^{00}=\dfrac{E_{1}E_{2}}{4 \pi r}t, \hspace{1 cm} T{}_{0i}^{\prime} \left( \partial^{2}\right) ^{-1}T{}^{0i}=-\dfrac{E_{1}E_{2}\sigma}{4 \pi r}t, \hspace{1 cm} T{}_{ij}^{\prime} \left( \partial^{2}\right) ^{-1}T{}^{ij}=\dfrac{E_{1}E_{2}\sigma^{2}}{4 \pi r}t,
\end{align}
which yield the potential energy between these two photons as 
\begin{align}
U= \dfrac{-2GE_{1}E_{2}}{ r} \left( 1-\sigma\right)^{2}.
\end{align}
So clearly  photons that move parallel  to each other do not see each other ($U=0$) at this level of approximation (namely the weak field limit and neglecting the spins). As noted this is also valid in mGR . Can this amusing result have observable consequences? Imagine two photons created by two infinitesimally close sources in a far away pulsar (or some other source), and travel side by side towards the earth for billions of years. If they interacted at all this would clearly have some non-trivial effects on their polarization. This discussion will be fully addressed elsewhere.  

\section{Spin-spin, spin-orbit, orbit-orbit interactions in massive gravity}

Consider two conserved sources, $\partial_{\mu}T^{\mu\nu}=0$, each having the following energy-momentum tensor components \cite{weinberg}

\begin{eqnarray} 
	T_{00}=T^{\left(0\right)}_{00}+T^{\left(2\right)}_{00}, \hspace{10 mm}
	T_{i0}=T^{\left(1\right)}_{i0}, \hspace{10 mm}
	T_{ij}=T^{\left(2\right)}_{ij} , 
\end{eqnarray}
where the relevant parts read
\begin{eqnarray} 
	T^{\left(0\right)}_{00}&=&m\delta\left(\vec{x}-\vec{x}_{a}\right),\nonumber \\
	T^{\left(2\right)}_{00}&=&\frac{1}{2}m\vec{v}^{2} \delta \left(\vec{x}-\vec{x}_{a}\right)-\frac{1}{2}J^{k}\,v^{i}\epsilon^{ikj}\partial_{j}\delta \left(\vec{x}
	-\vec{x}_{a}\right),\nonumber \\
	T^{\left(1\right)}_{i0}&=&-m v^{i} \delta \left(\vec{x}-\vec{x}_{a}\right)+\frac{1}{2}J^{k}\,\epsilon^{ikj}\partial_{j}\delta\left(\vec{x}
	-\vec{x}_{a}\right),\nonumber \\
	T^{\left(2\right)}_{ij}&=&mv^{i}v^{j}\delta \left(\vec{x}-\vec{x}_{a}\right)+J^{l}v^{(i}\epsilon^{j)kl}\partial_{k}\delta \left(\vec{x}
	-\vec{x}_{a}\right). 
	\label{en_mom}
\end{eqnarray}
Here $\vec{x}_{a}=\vec{x}_{a}(t)$ is the location of the particle and $v^{(i}\epsilon^{j)kl}$ refers to symmetrization with a $1/2$ factor. Note that our signature is $(-+++)$.  The relative signs and coefficients are fixed by  the requirement that at the first order (namely up to $O(v^{2})$ and $O(vJ)$ ) conservation equations are satisfied. Since it could sometimes be a little confusing let us explicitly compute two examples.
At the first order, 
\begin{align}
\partial_{0}T^{00}+\partial_{i}T^{i0}=m_{a}\partial_{0}\delta\left(\vec{x}-\vec{x}_{a}\right)+m_{a}v^{i}_{a}\partial_{i}\delta\left(\vec{x}-\vec{x}_{a}\right)-\frac{1}{2}J_{a}^{k}\,\epsilon^{ikj}\partial_{i}\partial_{j}\delta\left(\vec{x}
-\vec{x}_{a}\right)
\label{cons1}
\end{align}
The last term vanishes obviously, and we have $\partial_{0}\delta\left(\vec{x}-\vec{x}_{a}\right)=-v^{l}_{a}\partial_{l}\delta\left(\vec{x}-\vec{x}_{a}\right)$ since   $\vec{v}=\dfrac{d\vec{x}}{dt}$. Thus (\ref{cons1}) vanishes at the first order.
Similarly we have \begin{equation}
\partial_{0}T^{0j}+\partial_{i}T^{ij}=-m_{a}v^{j}_{a}v^{m}_{a}\partial_{m}\delta\left(\vec{x}-\vec{x}_{a}\right)+\frac{1}{2}J_{a}^{k}\,v^{m}_{a}\epsilon^{jkn}\partial_{n}\partial_{m}\delta\left(\vec{x}
-\vec{x}_{a}\right)+m_{a}v^{i}_{a}v^{j}_{a}\partial_{i}\delta\left(\vec{x}-\vec{x}_{a}\right)+\dfrac{1}{2}J_{a}^{l}v^{i}_{a}\epsilon^{jkl}\partial_{k}\partial_{i}\delta\left(\vec{x}
-\vec{x}_{a}\right)=0.
\end{equation}

In what follows, we will need the following identities
\begin{align}
\partial_{k}r= \dfrac{(x^{k}-x^{\prime k})}{r}=\hat{r}^{k}, \hspace{1cm} \partial_{k}\dfrac{1}{r}=\dfrac{-(x^{k}-x^{\prime k})}{r^{3}}=\dfrac{-\hat{r}^{k}}{r^{2}}, \nonumber \\
\partial_{k^{\prime}}r= \dfrac{-(x^{k}-x^{\prime k})}{r}=-\hat{r}^{k}, \hspace{1cm} \partial_{k^{\prime}}\dfrac{1}{r}=\dfrac{(x^{k}-x^{\prime k})}{r^{3}}=\dfrac{\hat{r}^{k}}{r^{2}}, \nonumber \\
\partial_{k}\partial_{n^{\prime}}r=\dfrac{1}{r}\left( -\delta^{kn}+\hat{r}^{k}\hat{r}^{n}\right) , \hspace{1cm}
\partial_{k}\partial_{n^{\prime}}\dfrac{1}{r}= \dfrac{1}{r^{3}}\left( \delta^{kn}-3\hat{r}^{k}\hat{r}^{n}\right),
\label{derivations}
\end{align}
where we have assumed, $r\neq0$  otherwise one picks up a "Fermi contact" term in the last expression. But this is irrelevant for our computation since we are in the long separation regime of two sources.

\subsection{Gravitomagnetic Effects in General Relativity}

In $D=3+1$, we have

\begin{equation}
	4Ut=-2\kappa T{}_{00}^{\prime}(\partial^{2})^{-1}T^{00} -4\kappa T{}_{0i}^{\prime}(\partial^{2})^{-1}T^{0i} 
	-2\kappa T{}_{ij}^{\prime}(\partial^{2})^{-1}T^{ij}+\kappa  T^{\prime}(\partial^{2})^{-1}T,
	\label{gr}
\end{equation}
where the trace of the energy-momentum tensor reads 
\begin{equation}
T=-T{}_{00}+\delta^{ij}T{}_{ij}=-m_{a}\delta\left(\vec{x}-\vec{x}_{a}\right)+\frac{1}{2}m_{a}\vec{v}^{2}_{a} \delta\left(\vec{x}-\vec{x}_{a}\right)-\frac{1}{2}J_{a}^{k}\,v^{i}_{a}\epsilon^{ikj}\partial_{j}\delta\left(\vec{x}
-\vec{x}_{a}\right).
\end{equation}

Let us compute the terms separately. The energy density terms read

\begin{align}
	T{}_{00}(\partial^{2})^{-1}T^{\prime 00}=&\bigg [  m_{1}\delta\left(\vec{x}-\vec{x}_{1}\right)+\frac{1}{2}m_{1}\vec{v}^{2}_{1} \delta\left(\vec{x}-\vec{x}_{1}\right) -\frac{1}{2}J_{1}^{l}\,v^{i}_{1}\epsilon^{ilk}\partial_{k}\delta\left(\vec{x}
	-\vec{x}_{1}\right)\bigg ]  (\partial^{2})^{-1} \nonumber \\
	&
	\bigg [ m_{2}\delta\left(\vec{x^{\prime}}-\vec{x}_{2}\right)+\frac{1}{2}m_{2}\vec{v}^{2}_{2} \delta\left(\vec{x^{\prime}}-\vec{x}_{2}\right)  -\frac{1}{2}J_{2}^{m}\,v^{j}_{2}\epsilon^{jmn}\partial_{n}^{\prime}\delta\left(\vec{x^{\prime}}
	-\vec{x}_{2}\right)\bigg ],
\end{align}
with the individual terms yielding the following contributions

\begin{equation}
	m_{1}\delta\left(\vec{x}-\vec{x}_{1}\right)(\partial^{2})^{-1}m_{2}\delta\left(\vec{x^{\prime}}-\vec{x}_{2}\right)=\frac{m_{1}m_{2}}{4\pi r}t,
\end{equation}

\begin{equation}
	m_{1}\delta\left(\vec{x}-\vec{x}_{1}\right)(\partial^{2})^{-1}\frac{1}{2}m_{2}\vec{v}^{2}_{2} \delta\left(\vec{x^{\prime}}-\vec{x}_{2}\right)=\frac{1}{2}\frac{m_{1}m_{2}\vec{v}^{2}_{2}}{4\pi r}t,
\end{equation}

\begin{equation}
	-\frac{1}{2}m_{1}\delta\left(\vec{x}-\vec{x}_{1}\right)(\partial^{2})^{-1}J_{2}^{m}\,v^{j}_{2}\epsilon^{jmn}\partial_{n}^{\prime}\delta\left(\vec{x^{\prime}}
	-\vec{x}_{2}\right) 
	=\frac{1}{2}
	\frac{m_{1}(\hat{r}\times \vec{v}_{2}).\vec{J_{2}}}{4\pi r^{2}}t,	
\end{equation}

\begin{equation}
\frac{1}{2}m_{1}\vec{v}^{2}_{1} \delta\left(\vec{x}-\vec{x}_{1}\right)(\partial^{2})^{-1}m_{2}\delta\left(\vec{x^{\prime}}-\vec{x}_{2}\right)=\frac{1}{2}\frac{m_{1}m_{2}\vec{v}^{2}_{1}}{4\pi r}t,
\end{equation}

\begin{equation}
	-\frac{1}{2}J_{1}^{l}\,v^{i}_{1}\epsilon^{ilk}\partial_{k}\delta\left(\vec{x}
	-\vec{x}_{1}\right) (\partial^{2})^{-1} m_{2}\delta\left(\vec{x^{\prime}}-\vec{x}_{2}\right)
	=-\frac{1}{2}
	\frac{m_{2}(\hat{r}\times \vec{v}_{1}).\vec{J_{1}}}{4\pi r^{2}}t.
\end{equation}

Observe that we have dropped the higher order term $O(J^{2}v^{2})$ to have a consistent expression. The collection of these terms yield 
\begin{align}
	-2\kappa T{}_{00}(\partial^{2})^{-1}T^{\prime 00}=-2\kappa \bigg [ \frac{m_{1}m_{2}}{4\pi r}\bigg (1+\frac{\vec{v}^{2}_{1}+\vec{v}^{2}_{2}}{2}\bigg )+\frac{1}{4\pi r^{2}}
	\bigg (\frac{m_{1}(\hat{r}\times \vec{v}_{2})\cdot\vec{J_{2}}}{2}-\frac{m_{2}(\hat{r}\times \vec{v}_{1} )\cdot\vec{J_{1}}}{2} \bigg )\bigg ]t.
\end{align}

Similarly the trace-trace interaction reads 

\begin{align}
	T^{\prime}(\partial^{2})^{-1}T= &\bigg [ -m_{1}\delta\left(\vec{x}-\vec{x}_{1}\right)+\frac{1}{2}m_{1}\vec{v}^{2}_{1} \delta\left(\vec{x}-\vec{x}_{1}\right) -\frac{1}{2}J_{1}^{l}\,v^{i}_{1}\epsilon^{ilk}\partial_{k}\delta\left(\vec{x}
	-\vec{x}_{1}\right)\bigg ] (\partial^{2})^{-1} \nonumber \\
	&\bigg [ -m_{2}\delta\left(\vec{x^{\prime}}-\vec{x}_{2}\right)+\frac{1}{2}m_{2}\vec{v}^{2}_{2} \delta\left(\vec{x^{\prime}}-\vec{x}_{2}\right)  -\frac{1}{2}J_{2}^{m}\,v^{j}_{2}\epsilon^{jmn}\partial_{n}^{\prime}\delta\left(\vec{x^{\prime}}
	-\vec{x}_{2}\right)\bigg ],
\end{align}
which after carrying out the relevant integrals becomes
\begin{align}
	\kappa T^{\prime}(\partial^{2})^{-1}T=	\kappa\bigg [\frac{m_{1}m_{2}}{4\pi r}\bigg(1+\frac{-\vec{v}^{2}_{1}-\vec{v}^{2}_{2}}{2}\bigg)
	+\frac{1}{4\pi r^{2}}\bigg (
	-\frac{m_{1}(\hat{r}\times \vec{v}_{2})\cdot\vec{J_{2}}}{2}+
	\frac{m_{2}(\hat{r}\times \vec{v}_{1})\cdot\vec{J_{1}}}{2} \bigg )\bigg ] t.
\end{align}
Observe that $T{}_{ij}^{\prime}(\partial^{2})^{-1}T{}^{ij}$  term in (\ref{gr}) yields higher order corrections which we drop, hence the final piece is the $T{}_{0i}^{\prime}(\partial^{2})^{-1}T^{0i}$ term which reads 

\begin{align}
	T{}_{0i}^{\prime}(\partial^{2})^{-1}T^{0i}=\bigg [ -m_{1}v^{i}_{1}\delta\left(\vec{x}-\vec{x}_{1}\right)+\frac{1}{2}J_{1}^{k}\,\epsilon^{ikj}\partial_{j}\delta\left(\vec{x}
	-\vec{x}_{1}\right)\bigg ](\partial^{2})^{-1}\bigg [ m_{2}v^{i}_{2}\delta\left(\vec{x^{\prime}}-\vec{x}_{2}\right)-\frac{1}{2}J_{2}^{l}\,\epsilon^{ilm}\partial^{\prime}_{m}\delta\left(\vec{x^{\prime}}
	-\vec{x}_{2}\right)\bigg ],
\end{align}
whose explicit computation gives 
\begin{equation}
-4\kappa T{}_{0i}^{\prime}(\partial^{2})^{-1}T^{0i}=-4\kappa \bigg [ -\dfrac{m_{1}m_{2}\vec{v}_{1}\cdot\vec{v}_{2} }{4\pi r}      +\dfrac{1}{4 \pi r^{2}} \dfrac{1}{2}   \bigg ( -m_{1}(\hat{r}\times \vec{v}_{1}  )\cdot \vec{J}_{2}+m_{2}(\hat{r}\times \vec{v}_{2}  )\cdot \vec{J}_{1}\bigg)  - \dfrac{1}{4 \pi r^{3}} \dfrac{2\vec{J}_{1}\cdot \vec{J}_{2}-3(\hat{r}\times \vec{J}_{1})\cdot(\hat{r}\times \vec{J}_{2})}{4}\bigg]t.
\end{equation}

Collecting all these parts above, the potential energy in GR at the desired order becomes

\begin{eqnarray}
U_{GR} &=& -\frac{G}{r}  m_{1}m_{2} \left [ 1+\frac{3}{2}\vec{v}^{2}_{1}+\frac{3}{2}\vec{v}^{2}_{2}-4\vec{v}_{1}\cdot \vec{v}_{2} \right ]
-\frac{G}{r^{3}}\left[\vec{J_{1}}\centerdot\vec{J_{2}}-3\vec{J_{1}}\centerdot\hat{r}\,
\vec{J_{2}}\centerdot\hat{r}\right] \nonumber \\
&&-\frac{G}{r^{2}}  \left [ \frac{3m_{1}(\hat{r}\times \vec{v}_{2})\cdot\vec{J_{2}}}{2}-\frac{3m_{2}(\hat{r}\times \vec{v}_{1})\cdot\vec{J_{1}}}{2}-2m_{1}(\hat{r}\times \vec{v}_{1})\cdot\vec{J_{2}}+2m_{2}(\hat{r}\times \vec{v}_{2})\cdot\vec{J_{1}}
	\right ], \label{grss}
\end{eqnarray}
which include the Newtonian potential energy, plus relativistic corrections such as spin-spin and spin-orbit effects. Note that $\vec{v}_{1}$ and $\vec{v}_{2}$ are defined with respect to a frame at rest.

\subsection{Gravitomagnetic Effects in Massive Gravity}
Analogous computation in massive GR is somewhat more involved but still it is straightforward,  we relegate to the Appendix the details and quote the final result here. One starts with 
\begin{align}
	4Ut= -2\kappa T{}_{00}^{\prime}\left\{ \partial^{2}-m_{g}^{2}\right\} ^{-1}T^{00} 
	+\frac{2\kappa}{3}T^{\prime}\left\{ \partial^{2}-m_{g}^{2}\right\} ^{-1}T 
	-4\kappa T{}_{0i}^{\prime}\left\{ \partial^{2}-m_{g}^{2}\right\} ^{-1}T^{0i}-2\kappa T{}_{ij}^{\prime}\left\{ \partial^{2}-m_{g}^{2}\right\} ^{-1}T^{ij} . \label{massive_pot}
\end{align}
Again the $T{}_{ij}^{\prime}-T{}^{ij}$ term will not contribute at the order we are working. At the desired order, gravitomagnetic potential in mGR is

	\begin{eqnarray}
	U_{mGR}&=&  -G e^{-m_{g}r} \frac{m_1 m_2}{r}\left [  \frac{4}{3}   +   \frac{4}{3}\vec{v}^{2}_{1}+\frac{4}{3}\vec{v}^{2}_{2} - 4 \vec{v}_{1} \cdot \vec{v}_{2}   \right] \nonumber \\
	&& -\frac{Ge^{-m_{g}r}\left(1+m_{g}r
		+m_{g}^{2}r^{2}\right)}{r^{3}}\left[\vec{J_{1}}\centerdot\vec{J_{2}}-3\vec{J_{1}}\centerdot\hat{r}\,
	\vec{J_{2}}\centerdot\hat{r}\frac{\left(1+m_{g}r+\frac{1}{3}m_{g}^{2}r^{2}\right)}{\left(1+m_{g}r+m_{g}^{2}r^{2}\right)} 
	\right]  \label{mgrss} \\
	&&
	-\frac{Ge^{-m_{g}r}}{r^{2}} \left(1+m_{g}r
	\right) \left[ \frac{4m_{1}\left( \hat{r}\times \vec{v}_{2}\right) \cdot \vec{J_{2}}}{3}-\frac{4m_{2}\left( \hat{r}\times \vec{v}_{1}\right) \cdot \vec{J_{1}}}{3}-2m_{1}\left( \hat{r}\times\vec{v}_{1}\right)\cdot \vec{J_{2}}	+2m_{2}\left( \hat{r}\times\vec{v}_{2}\right)\cdot \vec{J_{1}}\right] . \nonumber 
	\end{eqnarray}

In the $m_{g}\longrightarrow 0$ limit for not too large distances from (\ref{mgrss}), one obtains
	\begin{eqnarray}
		U_{mGR}&\Longrightarrow&  -G  \frac{m_1 m_2}{r} \left [  \frac{4}{3}  +   \frac{4}{3}\vec{v}^{2}_{1}+\frac{4}{3}\vec{v}^{2}_{2} - 4 \vec{v}_{1} \cdot \vec{v}_{2}   \right] \nonumber \\
		&& -\frac{G}{r^{3}}\left[\vec{J_{1}}\centerdot\vec{J_{2}}-3\vec{J_{1}}\centerdot\hat{r}\,
		\vec{J_{2}}\centerdot\hat{r}
		\right]  \\
		&&
		-\frac{G}{r^{2}}  \left[ \frac{4m_{1}\left( \hat{r}\times \vec{v}_{2}\right) \cdot \vec{J_{2}}}{3}-\frac{4m_{2}\left( \hat{r}\times \vec{v}_{1}\right) \cdot \vec{J_{1}}}{3}-2m_{1}\left( \hat{r}\times\vec{v}_{1}\right)\cdot \vec{J_{2}}	+2m_{2}\left( \hat{r}\times\vec{v}_{2}\right)\cdot \vec{J_{1}}\right] . \nonumber
	\end{eqnarray}
So the spin-spin part smoothly reduces to the GR expression in this limit, while a new discontinuity arises (a discrete $8/9$ difference between GR and mGR ) in the $O(v^{2})$ and $O(vJ)$ terms. Observe that, if one takes the large $r$ limit first, then goes to the  $m_{g}\longrightarrow 0$ limit, then, as noted in \cite{Gullu_Tekin1}, the spin-spin part also gives a distinctly different answer from GR. As discussed in the Introduction, a detailed analysis in that work revealed that the discrete difference arises for distances $m_{g}\,r\geq 1.62$.
\section{gravitomagnetic effects in quadratic gravity}
The relevant potential energy is 
\begin{align}
	U_{quad} t  =-\frac{\kappa}{2}T{}_{\mu\nu}^{\prime}\left(\partial^{2}\right)^{-1}T^{\mu\nu}+\frac{\kappa T^{\prime}
		\left(\partial^{2}\right)^{-1}T}{4}
	 +\frac{\kappa}{2}T{}_{\mu\nu}^{\prime}\left(\partial^{2}-m_{\beta}^{2}\right)^{-1}T^{\mu\nu}-\frac{\kappa T^{\prime}
		\left(\partial^{2}-m_{\beta}^{2}\right)^{-1}T}{6}
	 -\frac{\kappa T^{\prime}\left(\partial^{2}-m_{c}^{2}\right)^{-1}T}{12},
\end{align}
where there are two additional massive modes: a massive spin-2 graviton with $m_{\beta}^{2}=-\frac{1}{\kappa\beta}$ and a massive spin-0 mode with $m_{c}^{2}=\frac{1}{4\kappa\left(3\alpha+\beta\right)}$.
Massless spin-2 mode of GR is intact and so there will be terms added to (\ref{grss}). Here we define  $U_{quad}\equiv U_{GR}+U_{2}$, where
\begin{align}
	U_{2} t & =
	\frac{\kappa}{2}T{}_{00}^{\prime}\left(\partial^{2}-m_{\beta}^{2}\right)^{-1}T^{00}+\kappa T{}_{0i}^{\prime}\left(\partial^{2}-m_{\beta}^{2}\right)^{-1}T^{0i}+\frac{\kappa}{2} T{}_{ij}^{\prime}\left(\partial^{2}-m_{\beta}^{2}\right)^{-1}T^{ij}-\frac{\kappa }{6}T^{\prime}
	\left(\partial^{2}-m_{\beta}^{2}\right)^{-1}T\nonumber \\
	& -\frac{\kappa }{12}T^{\prime}\left(\partial^{2}-m_{c}^{2}\right)^{-1}T.
	\label{quad_amplitude}
\end{align}
The ensuing computation is similar so we can simply write the final result
\begin{align}
U_{2}= \dfrac{Gm_{1}m_{2}}{r}\left[ \left(  \dfrac{4}{3}+\dfrac{7}{3}\vec{v_{1}}^{2}+\dfrac{7}{3}\vec{v_{2}}^{2} \right) e^{-rm_{\beta}} -\left( \dfrac{1}{3}-\dfrac{1}{6}\vec{v_{1}}^{2}-\dfrac{1}{6}\vec{v_{2}}^{2} \right)e^{-rm_{c}} \right] \nonumber \\
+\dfrac{G}{r^{2}} \bigg [ \left( \dfrac{4}{3} m_{1}(\hat{r}\times\vec{v_{2}})\cdot \vec{J_{2}}-\dfrac{4}{3} m_{2}(\hat{r}\times\vec{v_{1}})\cdot \vec{J_{1}} -2m_{1}(\hat{r}\times\vec{v_{1}})\cdot \vec{J_{2}}+2m_{2}(\hat{r}\times\vec{v_{2}})\cdot \vec{J_{1}} \right) \left( 1+rm_{\beta}\right) e^{-rm_{\beta}} \nonumber \\
+ \left( \dfrac{1}{6}m_{1}(\hat{r}\times\vec{v_{2}})\cdot \vec{J_{2}}-\dfrac{1}{6}m_{2}(\hat{r}\times\vec{v_{1}})\cdot \vec{J_{1}} \right) \left( 1+rm_{c}\right) e^{-rm_{c}} \bigg ]\nonumber \\
+\dfrac{G}{r^{3}} \left(1+rm_{\beta}+r^{2}m_{\beta}^{2} \right) \bigg [ \vec{J_{1}}\cdot \vec{J_{2}}-3\vec{J_{1}}\cdot\hat{r} \vec{J_{2}}\cdot\hat{r} \dfrac{\left(1+rm_{\beta}+\frac{1}{3}r^{2}m_{\beta}^{2} \right) }{\left(1+rm_{\beta}+r^{2}m_{\beta}^{2} \right)}   \bigg ] e^{-rm_{\beta}}.
\end{align}

Let us compute the $r\longrightarrow 0$ limit, where we expect higher curvature terms to play a role, in the potential 
$U_{quad}$. In this limit we obtain 
\begin{equation}
U_{quad} \xrightarrow{r\longrightarrow 0} \dfrac{Gm_{1}m_{2}}{r} (v^{2}_{1}+v^{2}_{2}) + \mbox{constant}.
\end{equation} 
Observe that all the spin-spin, spin-orbit terms dropped as the quadratic parts cancel the GR parts but a repulsive $O(v^{2})$ term survives. Repulsive nature of quadratic gravity at small separations is expected and in fact, this is the reason why the theory is less divergent in the UV regime compared to Einstein's theory. But it is well-known that non-zero $\beta$ gives a spin-2 ghost \cite{stelle}. 

\section{conclusions}
In our recent work \cite{Gullu_Tekin1}, we initiated a study of gravitomagnetic effects in Fierz-Pauli theory (mGR) which is the \emph{unique} viable linear massive gravity that any non-linear version must reduce at the weak field limit (see \cite{paw} for related issues). In that work we computed the spin-spin interaction between distant sources (such as two galaxy clusters, or galaxies) and found that GR and mGR yield different prediction. In the current work, we extended these discussions to the case where the sources also have velocities and orbital motion as well as spins. Our computation is based on the tree-level graviton exchange diagram and we kept the terms up to and including $O(v^{2})$, $O(vJ)$, $O(J^{2})$.  mGR has distinctly different predictions compared to GR in all these orders, extending the well-known vdVZ result in the Newtonian potential energy. What is quite interesting is that, especially for the spin-spin interaction part the distinct differences arise for separations that satisfy $m_{g}\,r\geq 1.62$.  It is important to realize that for these distances, we are well outside the Vanishtein radius  \cite{Vain} and so the linearized gravity is perfectly legitimate. If one considers the Compton wavelength of the graviton to be at the same order as the Hubble radius of the universe, $\lambda_{c}=\frac{2\pi}{m_{g}}\simeq R_{u}\simeq 4.4 \times 10^{26} m $ \cite{goldh}, then massive gravity shows its most dramatic effects just at the edge of the observable universe.

Our computations suggest that in addition to the expectation that  graviton mass can explain the accelerated expansion of the universe, there are further, possibly observable effects of massive gravity on the spin orientations of galaxy clusters, or galaxies. It is very early to actually quote data on this but recent observations of unexplained spin alignments of 19 pulsars separated for billions of light years \cite{hutse} could possibly be due to massive gravity. See also the spin alignment observations of galaxies in \cite{longo}, \cite{shamir}.

We have also calculated the gravitomagnetic effects in Einstein + quadratic gravity theories, extending the discussion of \cite{Gullu_Tekin1} and found that as $r\longrightarrow 0 $, the potential energy has a repulsive part due to the kinetic energy of sources while the Newtonian, spin-spin and spin-orbit interactions vanish.

Finally, for weak fields and slow velocities, at the lowest order, GR has a nice formulation which resembles Maxwell's theory, where the electric field is  replaced by the gravitoelectric field and the magnetic field is replaced by the  gravitomagnetic field, the electric charge density is the replaced by mass density and the electric current is replaced by the mass current.

For the massive case, the situation becomes more complicated since both in the massive Maxwell's theory that is the Proca theory and massive gravity, $A_{\mu}$ and  $h_{\mu \nu}$ become "physical" hence they must appear in the field equations.

\subsection{Acknowledgment}

B.T. is partially supported by T\"{U}B\.{I}TAK grant 113F155, and thanks to A. Zee for useful discussion on light by light interactions in GR.

\section{appendix: Details of the computations}
Here we give the details of the computation for the massive gravity case. We start with (\ref{massive_pot}) and integrate the terms separately. The energy-density interaction part reads
 \begin{align}
T{}_{00}^{\prime}\left(  \partial^{2}-m_{g}^{2}\right)  ^{-1}T^{00}=\bigg [ m_{1}\delta\left(\vec{x}-\vec{x}_{1}\right)+\frac{1}{2}m_{1}\vec{v}^{2}_{1} \delta\left(\vec{x}-\vec{x}_{1}\right)  -\frac{1}{2}J_{1}^{l}\,v^{i}_{1}\epsilon^{ilk}\partial_{k}\delta\left(\vec{x}
-\vec{x}_{1}\right)\bigg ] \left(  \partial^{2}-m_{g}^{2}\right)  ^{-1} \nonumber \\
\bigg [ m_{2}\delta\left(\vec{x^{\prime}}-\vec{x}_{2}\right)+\frac{1}{2}m_{2}\vec{v}^{2}_{2} \delta\left(\vec{x^{\prime}}-\vec{x}_{2}\right) 
-\frac{1}{2}J_{2}^{m}\,v^{j}_{2}\epsilon^{jmn}\partial_{n}^{\prime}\delta\left(\vec{x^{\prime}}
-\vec{x}_{2}\right)\bigg ], \nonumber \\
=t\bigg [ m_{1}m_{2}\left(1+\frac{\vec{v}^{2}_{1}+\vec{v}^{2}_{2}}{2}+\dfrac{\vec{v}^{2}_{1}\vec{v}^{2}_{2}}{4}\right)\times\frac{1}{\left(2\pi\right)^{\frac{3}{2}}}\left[\left(\frac{m_{g}}{r}\right)^{\frac{1}{2}}K_{\frac{1}{2}}\left(r\, m_{g}\right) \right] \bigg ]\nonumber \\ +t \bigg [\dfrac{1}{2}\left(m_{1} J_{2}^{m}\,v^{j}_{2}\epsilon^{jmn}\partial_{n}^{\prime}+m_{2}J_{1}^{l}\,v^{i}_{1}\epsilon^{ilk}\partial_{k}\right) \nonumber \\
+\dfrac{1}{4}\left(m_{1}\vec{v}^{2}_{1}J_{2}^{m}\,v^{j}_{2}\epsilon^{jmn}\partial_{n}^{\prime}+m_{2}\vec{v}^{2}_{2} J_{1}^{l}\,v^{i}_{1}\epsilon^{ilk}\partial_{k}  + J_{1}^{l}\,v^{i}_{1}\epsilon^{ilk} J_{2}^{m}\,v^{j}_{2}\epsilon^{jmn} \partial_{k}\partial_{n}^{\prime}   \right) \bigg ] \nonumber \\
\times\frac{1}{\left(2\pi\right)^{\frac{3}{2}}}\left[\left(\frac{m_{g}}{r}\right)^{\frac{1}{2}}K_{\frac{1}{2}}\left(r\, m_{g}\right) \right] ,
\label{app1}
\end{align}
where the Bessel functions in three dimensions are
\begin{align}
	K_{\frac{1}{2}}\left(r\, m_{x}\right)=\frac{e^{-r\, m_{x}}}{\sqrt{r\, m_{x}}} \sqrt{\frac{\pi}{2}}, 
	K_{\frac{3}{2}}\left(r\, m_{x}\right)=\frac{e^{-r\, m_{x}}}{\sqrt{r\, m_{x}}} \sqrt{\frac{\pi}{2}} \left( 1+\frac{1}{r\, m_{x}}\right), 
	K_{\frac{5}{2}}\left(r\, m_{x}\right)=\frac{e^{-r\, m_{x}}}{\sqrt{r\, m_{x}}} \sqrt{\frac{\pi}{2}} \left( 1+\frac{3}{r\, m_{x}}+\frac{3}{(r\, m_{x})^{2}}\right) .
\end{align}
Now using $\partial_{z}(z^{-\nu}K_{\nu}\left(z\right))=-z^{-\nu}K_{\nu +1}\left(z\right)$ , we have

\begin{align}
\partial_{k}^{\prime} \left[ (r\, m_{g})^{-\nu}K_{\nu}\left(r\, m_{g}\right)\right] =\frac{(x^{k}-x^{\prime k})}{r}\, m_{g}\left(r\, m_{g}\right)^{-\nu}K_{\nu +1}\left(r\, m_{g}\right), \nonumber \\
\partial_{k} \left[ (r\, m_{g})^{-\nu}K_{\nu}\left(r\, m_{g}\right)\right] =-\frac{(x^{k}-x^{\prime k})}{r}\, m_{g}\left(r\, m_{g}\right)^{-\nu}K_{\nu +1}\left(r\, m_{g}\right), \nonumber \\  
	 \partial_{k} \partial_{n}^{\prime}\left[ (r\, m_{g})^{-\nu}K_{\nu}\left(r\, m_{g}\right)\right] = m_{g}^{2}\left[ \delta^{kn}\left(r\, m_{g}\right)^{-(\nu+1)}K_{\nu +1}\left(r\, m_{g}\right) -\hat{r}^{k}\hat{r}^{n}\left(r\, m_{g}\right)^{-\nu}K_{\nu +2}\left(r\, m_{g}\right)\right].
	   \end{align}
In order to simplify the computation, we write (\ref{app1}) as 
\begin{align}
T{}_{00}^{\prime}\left\{ \partial^{2}-m_{g}^{2}\right\} ^{-1}T^{00}=\frac{m_{g}}{\left(2\pi\right)^{\frac{3}{2}}} t \bigg \{ m_{1}m_{2}\left(1+\frac{\vec{v}^{2}_{1}+\vec{v}^{2}_{2}}{2}+\dfrac{\vec{v}^{2}_{1}\vec{v}^{2}_{2}}{4}\right)\bigg \} \times\left[\left(r\, m_{g}\right)^{\frac{-1}{2}}K_{\frac{1}{2}}\left(r\, m_{g}\right) \right] \nonumber \\  +\frac{m_{g}}{\left(2\pi\right)^{\frac{3}{2}}} t\bigg \{\dfrac{1}{2}\left(m_{1} J_{2}^{m}\,v^{j}_{2}\epsilon^{jmn}\partial_{n}^{\prime}+m_{2}J_{1}^{l}\,v^{i}_{1}\epsilon^{ilk}\partial_{k}\right) \nonumber \\
+\dfrac{1}{4}\left(m_{1}\vec{v}^{2}_{1}J_{2}^{m}\,v^{j}_{2}\epsilon^{jmn}\partial_{n}^{\prime}+m_{2}\vec{v}^{2}_{2} J_{1}^{l}\,v^{i}_{1}\epsilon^{ilk}\partial_{k}  + J_{1}^{l}\,v^{i}_{1}\epsilon^{ilk} J_{2}^{m}\,v^{j}_{2}\epsilon^{jmn} \partial_{k}\partial_{n}^{\prime}   \right) \bigg \} \nonumber \\
\times\left[\left(r\, m_{g}\right)^{\frac{-1}{2}}K_{\frac{1}{2}}\left(r\, m_{g}\right) \right] .
\end{align}
After carrying out the integrations, one arrives at 
\begin{align}
m_{1} J_{2}^{m}\,v^{j}_{2}\epsilon^{jmn}\partial_{n}^{\prime}\left[\left(r\, m_{g}\right)^{\frac{-1}{2}}K_{\frac{1}{2}}\left(r\, m_{g}\right) \right]=m_{g} m_{1}(\hat{r}\times \vec{v}_{2})\cdot\vec{J_{2}}\left(r\, m_{g}\right)^{\frac{-1}{2}}K_{\frac{3}{2}}\left(r\, m_{g}\right),
\end{align}

\begin{align}
m_{2}J_{1}^{l}\,v^{i}_{1}\epsilon^{ilk}\partial_{k}\left[\left(r\, m_{g}\right)^{\frac{-1}{2}}K_{\frac{1}{2}}\left(r\, m_{g}\right) \right] =-m_{g} m_{2}(\hat{r}\times \vec{v}_{1})\cdot\vec{J_{1}}\left(r\, m_{g}\right)^{\frac{-1}{2}}K_{\frac{3}{2}}\left(r\, m_{g}\right),
\end{align}

\begin{align}
m_{1}\vec{v}^{2}_{1}J_{2}^{m}\,v^{j}_{2}\epsilon^{jmn}\partial_{n}^{\prime}\left[\left(r\, m_{g}\right)^{\frac{-1}{2}}K_{\frac{1}{2}}\left(r\, m_{g}\right) \right]=m_{g}m_{1}\vec{v}^{2}_{1}(\hat{r}\times \vec{v}_{2})\cdot\vec{J_{2}}\left(r\, m_{g}\right)^{\frac{-1}{2}}K_{\frac{3}{2}}\left(r\, m_{g}\right),
\end{align}

\begin{align}
m_{2}\vec{v}^{2}_{2} J_{1}^{l}\,v^{i}_{1}\epsilon^{ilk}\partial_{k}\left[\left(r\, m_{g}\right)^{\frac{-1}{2}}K_{\frac{1}{2}}\left(r\, m_{g}\right) \right]=-m_{g}m_{2}\vec{v}^{2}_{2}(\hat{r}\times \vec{v}_{1})\cdot\vec{J_{1}}\left(r\, m_{g}\right)^{\frac{-1}{2}}K_{\frac{3}{2}}\left(r\, m_{g}\right),
\end{align}

\begin{align}
J_{1}^{l}\,v^{i}_{1}\epsilon^{ilk} J_{2}^{m}\,v^{j}_{2}\epsilon^{jmn} \partial_{k}\partial_{n}^{\prime}\left[\left(r\, m_{g}\right)^{\frac{-1}{2}}K_{\frac{1}{2}}\left(r\, m_{g}\right) \right]=m^{2}_{g}\bigg [ \left[ \vec{v}_{1}.\vec{v}_{2}\vec{J}_{1}.\vec{J}_{2}-\vec{v}_{1}.\vec{J}_{2}\vec{v}_{2}.\vec{J}_{1} \right] \left(r\, m_{g}\right)^{\frac{-3}{2}}K_{\frac{3}{2}}\left(r\, m_{g}\right) \nonumber \\ 
-(\hat{r}\times \vec{v}_{1}).\vec{J_{1}}(\hat{r}\times \vec{v}_{2}).\vec{J_{2}}\left(r\, m_{g}\right)^{\frac{-1}{2}}K_{\frac{5}{2}}\left(r\, m_{g}\right) \bigg ].
\end{align}
Collection of these terms gives

\begin{align}
-\dfrac{\kappa}{2}T{}_{00}^{\prime}\left(  \partial^{2}-m_{g}^{2}\right)  ^{-1}T^{00}=-\kappa \frac{1}{2}\frac{m_{g}}{\left(2\pi\right)^{\frac{3}{2}}} \bigg [  m_{1}m_{2}\left(1+\frac{\vec{v}^{2}_{1}+\vec{v}^{2}_{2}}{2}\right)\left(r\, m_{g}\right)^{\frac{-1}{2}}K_{\frac{1}{2}}\left(r\, m_{g}\right) 
\nonumber \\ +\frac{1}{2}m_{g}\bigg ( m_{1}(\hat{r}\times \vec{v}_{2})\cdot\vec{J_{2}}-m_{2}(\hat{r}\times \vec{v}_{1})\cdot\vec{J_{1}}\bigg ) \left(r\, m_{g}\right)^{\frac{-1}{2}}K_{\frac{3}{2}}\left(r\, m_{g}\right)
\bigg ].
\end{align}
Similarly one finds

\begin{align}
\dfrac{\kappa}{6}T^{\prime}\left(  \partial^{2}-m_{g}^{2}\right)  ^{-1}T=\kappa \frac{1}{6}\frac{m_{g}}{\left(2\pi\right)^{\frac{3}{2}}}\bigg [ m_{1}m_{2} \bigg ( 1-\frac{1}{2}\vec{v}^{2}_{1}-\frac{1}{2}\vec{v}^{2}_{2}\bigg ) \left(r\, m_{g}\right)^{\frac{-1}{2}}K_{\frac{1}{2}}\left(r\, m_{g}\right) \nonumber \\ 
+m_{g} \bigg ( - \frac{m_{1}(\hat{r}\times \vec{v}_{2})\cdot\vec{J_{2}}}{2}+\frac{m_{2}(\hat{r}\times \vec{v}_{1})\cdot\vec{J_{1}}}{2}\bigg ) \left(r\, m_{g}\right)^{\frac{-1}{2}}K_{\frac{3}{2}}\left(r\, m_{g}\right)\bigg ],
\end{align}

\begin{align}
-\kappa T{}_{0i}^{\prime}\left(  \partial^{2}-m_{g}^{2}\right)  ^{-1}T^{0i}=-\kappa\frac{m_{g}}{\left(2\pi\right)^{\frac{3}{2}}} \bigg [ -m_{1}m_{2}\vec{v}_{1}\cdot\vec{v}_{2} \left(r\, m_{g}\right)^{\frac{-1}{2}}K_{\frac{1}{2}}\left(r\, m_{g}\right) \nonumber \\ -m_{g}\bigg ( \frac{m_{1}(\hat{r}\times \vec{v_{1}})\cdot\vec{J_{2}}}{2}-\frac{m_{2}(\hat{r}\times \vec{v_{2}})\cdot\vec{J_{1}}}{2}+m_{g}\frac{\vec{J_{1}} \cdot\vec{J_{2}}}{2} \left(r\, m_{g}\right)^{-1}\bigg ) \left(r\, m_{g}\right)^{\frac{-1}{2}}K_{\frac{3}{2}}\left(r\, m_{g}\right) \nonumber \\
+ \frac{1}{4}m_{g}^{2}
(\hat{r}\times \vec{J_{1}})\cdot(\hat{r}\times \vec{J_{2}}) \left(r\, m_{g}\right)^{\frac{-1}{2}}K_{\frac{5}{2}}\left(r\, m_{g}\right)            \bigg ].
\end{align}
Collecting all these pieces, one arrives at (\ref{mgrss}).

\end{document}